\documentclass{elsart}

\newcommand{\nc}{\newcommand}
\nc{\non}{\nonumber} 
\newcommand{\beq}{\begin{equation}}
\newcommand{\eeq}{\end{equation}}
\newcommand{\bega}{\begin{eqnarray}}
\newcommand{\ega}{\end{eqnarray}}

\usepackage{psfrag}

 \usepackage{graphicx}
\usepackage{amssymb}

\begin{document}
 
\begin{frontmatter}

% Title, authors and addresses

% use the thanksref command within \title, \author or \address for footnotes;
% use the corauthref command within \author for corresponding author footnotes;
% use the ead command for the email address,
% and the form \ead[url] for the home page:
% \title{Title\thanksref{label1}}
% \thanks[label1]{}
% \author{Name\corauthref{cor1}\thanksref{label2}}
% \ead{email address}
% \ead[url]{home page}
% \thanks[label2]{}
% \corauth[cor1]{}
% \address{Address\thanksref{label3}}
% \thanks[label3]{}

\title{Irregular diffusion \\ in the bouncing ball billiard} 

%\title{Diffusion of particles bouncing on a one-dimensional 
%vibrating corrugated floor}

% use optional labels to link authors explicitly to addresses:
% \author[label1,label2]{}
% \address[label1]{}
% \address[label2]{}

\author{L. M\'aty\'as and} 
\author{R. Klages} 
\address{Max Planck Institute for the Physics of Complex Systems, \\
         N\"othnitzer 38, 01187 Dresden, Germany}

\begin{abstract}
% Text of abstract
We call a system {\em bouncing ball billiard} if it consists of a particle
that is subject to a constant vertical force and bounces inelastically on a
one-dimensional vibrating periodically corrugated floor. Here we choose
circular scatterers that are very shallow, hence this billiard is a
deterministic diffusive version of the well-known bouncing ball problem on a
flat vibrating plate. Computer simulations show that the diffusion coefficient
of this system is a highly irregular function of the vibration frequency
exhibiting pronounced maxima whenever there are resonances between the
vibration frequency and the average time of flight of a particle. In addition,
there exist irregularities on finer scales that are due to higher-order
dynamical correlations pointing towards a fractal structure of this curve. We
analyze the diffusive dynamics by classifying the attracting sets and by
working out a simple random walk approximation for diffusion, which is
systematically refined by using a Green-Kubo formula.
\end{abstract}

\end{frontmatter}

% main text
%\section{}
%\label{}

\section{Introduction} 

An important challenge in the field of nonequilibrium statistical mechanics
is to achieve a more profound understanding of transport processes by starting
from the complete microscopic, deterministic, typically chaotic equations of
motion of a many-particle system \cite{Gas98,Do99,Vol02}. However, up to now
such a detailed analysis concerning the microscopic origin of nonequilibrium
transport could only be performed for certain classes of toy models. Very
popular among these models are systems that are low-dimensional, spatially
periodic and consist of a gas of moving point particles that do not interact
with each other but only with fixed scatterers.

For these chaotic dynamical systems it was found that, typically, the
respective transport coefficients are highly irregular functions of the
control parameter. These irregularities were exactly calculated for diffusion
in a simple abstract piecewise linear map \cite{KlDo95,Kla96,Kla99} and were
shown to be related to topological instabilities under parameter
variation. The same properties are exhibited by diffusive Hamiltonian particle
billiards such as the periodic Lorentz gas
\cite{KlDe00,KlKo02}, the flower-shaped billiard \cite{HKG02}, and for a
particle moving on a one-dimensional periodically corrugated floor
\cite{HaGa01}. Similar irregularities occur in nonhyperbolic 
maps with anomalous diffusion \cite{KoKl02L} in simple models exhibiting
(electric) currents \cite{GK01,LlNiRoMo95} as well as in the parameter
dependence of chemical reaction rates in reaction-diffusion processes
\cite{GaKl98}.  In most of the cases mentioned above, the deterministic
transport coefficients clearly showed fractal structures
\cite{KlDo95,Kla99,KlKo02,HaGa01,KoKl02L,GK01,GaKl98}. 
Seemingly analogous transport
anomalies were measured in experiments \cite{Weis91,Weiss} or have been
reported for models that appear to be rather close to experiments
\cite{WA01,TKK02}. However, still an experimental verification of the
fractal nature of irregular transport coefficients remains an open question.

In this paper we wish to contribute to this problem by introducing and
investigating a dynamical system that, we believe, is rather close to specific
experiments. For this purpose we consider a generalized version of the
interesting model described in Ref.\ \cite{HaGa01}, where a particle subject
to a constant vertical acceleration makes jumps on a periodically corrugated
floor. We generalize this model by including some friction at the collisions
and compensate this energy loss by periodic oscillations of the corrugated
floor. This generalized model is an example of what we call a {\em bouncing
ball billiard}. It is designed to be very close to the experiments on
diffusion in granular material performed in Refs.\
\cite{FaTeVuVi99,LoCoGo99,PrEgUr02}, thus we hope that our theoretical
predicitions can be verified by respective experiments. However, we emphasize
the fact that, in contrast to most experiments on granular
material,\footnote{An exception is the experiment performed in Ref.\
\cite{FaTeVuVi99} where the dynamics is exclusively due to the collisions of a
single particle with a corrugated surface.} here we study a granular gas that
consists of statistical ensembles of a one-particle system only. One should
therefore be very careful in concluding anything from the physics we discuss
here for the case of highly interacting many-particle systems.

The corrugated floor of our model is formed by circular scatterers that are
deliberately very shallow. Hence, another important limiting case of our model
is the famous {\em bouncing ball problem}, where an inelastic particle bounces
vertically on a flat vibrating surface. This problem has both been studied
experimentally \cite{Pi83,Pi85a,Pi85b,Pi88,Tu90} as well as theoretically
\cite{GuHo83,LL83,LuMe90,LuMe93}. It is well-known that the nonlinear dynamics
of the bouncing ball is very complex exhibiting one or more coexisting
attractors depending on the frequency of the floor \cite{FeGrHuYo96}, that is,
the system displays both ergodic and nonergodic dynamics. The main theme of
this paper will be to show that dynamical correlations in terms of phase
locking, as associated to these different attractors, have a huge impact on
diffusion in the bouncing ball billiard and that these regimes lead to strong
irregularities in the parameter-dependent diffusion coefficient on large and
small scales.

The paper is composed of six sections: In Section \ref{sec:BOBA} we outline
some important features of the bouncing ball problem. This knowledge provides
a roadmap in order to understand the diffusive behavior in the spatially
extended system. In Section \ref{sec:DEFOFMODEL} we give the full equations of
motion of the bouncing ball billiard and briefly explain the specific choice
of the control parameters. In Section \ref{sec:DIFCOEF} we present results for
the diffusion coefficient and for the average energy of the system as 
functions of the frequency of the vibrating floor. In Section
\ref{sec:PHASEDIAGRAM} we analyze the irregularities of the
parameter-dependent diffusion coefficient in full detail by relating them to
different dynamical regimes, as represented by certain structures in the
corresponding attracting sets. In Section \ref{sec:GRKU} we work out a simple
random walk approximation somewhat explaining features of the diffusion
coefficient on the coarsest scales. This approach is then refined by including
higher-order dynamical correlations based on a Green-Kubo formula for
diffusion thus providing detailed evidence for the existence of irregularities
on finer and finer scales. Section \ref{sec:CONCLUSIONS} contains a summary of
the results, an outline of further links to previous works, and an outlook
concerning further studies in this direction. In particular, we are trying to
encourage an {\em experimental realization} of our bouncing ball billiard.

\section{The bouncing ball problem: Arnold tongues} \label{sec:BOBA}

A standard problem of chaotic dynamics that was extensively studied is the one
of a ball subject to an external field that bounces inelastically on a flat
vibrating surface 
\cite{Pi83,Pi85a,Pi85b,Pi88,Tu90,GuHo83,LL83,LuMe90,LuMe93,Dev94}. At first
view this model appears to be rather simple, however, its equations of motion
are in fact highly nonlinear and generally do not allow any exact
solution. Hence, a good part of the investigations deals with simplified
versions of this model such as the so-called high bounce approximation leading
to the dissipative standard map \cite{Tu90,FeGrHuYo96}. Here we restrict to
the exact model only as nicely analyzed in Refs.\ \cite{Tu90,LuMe93} and
recall some important features that will be needed to understand the
forthcoming results.

Let us assume that the flat surface exhibits a sinusoidal motion
$y=-A\sin(\omega t)$, where $A$ and $\omega$ are the amplitude respectively
the frequency of the vibration. Between the bounces on the surface the ball
moves in a gravitational field with acceleration $g$.  If the mass of the
ground plate is assumed to be much larger than the mass of the ball, the
latter becomes a trivial quantity in the equations of motion that, for
convenience, we set equal to one. The bouncing ball problem then has four
control parameters: $A$, $\omega$, the gravitational acceleration $g$ and the
vertical restitution coefficient $\alpha$.  After a proper transformation of
the equations of motion $A$, $\omega$ and $g$ can be grouped into $\Gamma
\equiv A \omega^2/g$.  A particular complexity of the bouncing ball system is
related to the fact that for given values of the parameters $\Gamma$ and
$\alpha$ the system may posess more than one attractor, each with a different
basin of attraction. This is drastically exemplified in a study of the
high-bounce approximation revealing the coexistence of an arbitrarily large
number of attractors if the damping is small enough \cite{FeGrHuYo96}.

%***********************************************
\begin{figure}[h!]
{\hfill
\psfrag{alpha}{{\Huge $\alpha$}}
\psfrag{gamma}{{\Huge $\Gamma$}}
\scalebox{0.4}{\rotatebox{270}{\includegraphics{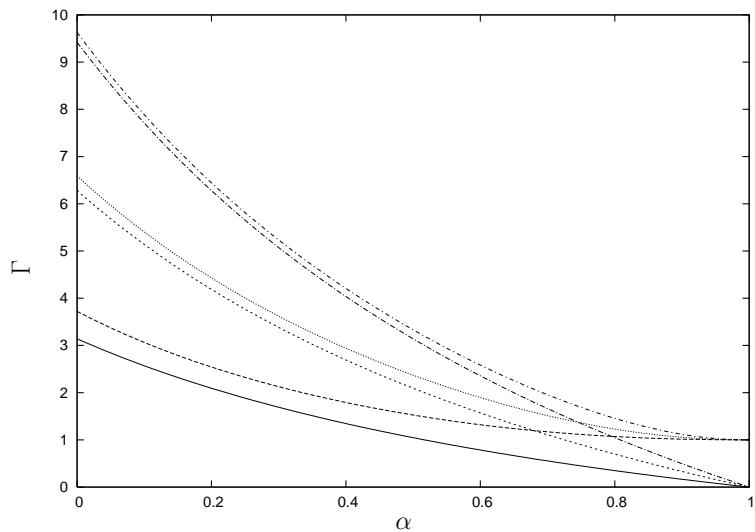}}}
\hfill}
\caption[small]{
The three major Arnold tongues of the bouncing ball problem in
parameter space. The adjacent odd and even lines represent the boundaries of
the resonances between the period of the driving $T_{\rm vibr}$ and the time
of flight between two bounces $T_{\rm flight}$, see Eq.\ (\ref{eq:artg}). 
From bottom to top it is $T_{\rm flight}/T_{\rm vibr}=1,2,3$. 
$\alpha$ and $\Gamma$
are dimensionless quantities.} 
\label{fig:at}
\end{figure}
%***********************************************

We will now focus on the Arnold tongues of the bouncing ball defining a series
of stable resonances. The largest ones were analyzed approximately in Refs.\
\cite{Pi85a,Pi88} and were lateron calculated exactly in Ref.\
\cite{LuMe93}.  An account of the hierarchy of smaller ones in the limit of
zero friction was approximately given in Ref.\ \cite{LuMe90}. A resonances of
order $k/1$ means that the period of the particle's flight between two
collisions $T_{\rm flight}$ is $k$ times larger than the period of the floor's
motion, $T_{\rm flight}=k\,T_{\rm vibr}\equiv k\,2\pi/\omega$.  The region in
parameter space $(\alpha,\Gamma)$ where a resonance of order $k$ exists can
then be calculated via linear stability analysis of the equations of motion to
\cite{LuMe93}
\beq
\pi \frac{1-\alpha}{1+\alpha}k < \Gamma <
\pi\left\{ \left[\frac{1-\alpha}{1+\alpha}k \right]^2 + 
\left[ \frac{2(1+\alpha^2)}{\pi(1+\alpha)^2} \right]^2\right\}^{1/2}\:.
\label{eq:artg}  
\eeq
These different major tongues are shown in Fig. \ref{fig:at}. Note that,
somewhat in contrast to simple Arnold tongues, here not all the inital
conditions in phase space may exhibit phase locking. That is, typically there
are in addition nonresonant trajectories that will ``stick'' to the surface
interrupted by sequences in which they are relaunched exhibiting jumps of
smaller and smaller amplitude. This particularly complicated dynamics was
denoted as ``self-reanimating chaos'' \cite{Pi88} or as ``chattering''
\cite{LuMe93}. In Ref.\ \cite{LuMe93}, a careful analysis indeed led to the
result ``that generic trajectories starting under experimental conditions
terminate in a region of chattering'', which seemed to be at variance with the
experimental observation of a Feigenbaum-like bifurcation scenario and a
respective transition to chaos \cite{Pi83,Pi85a,Pi85b,Pi88,Tu90}.
 
%%*****************************************************************
%%*****************************************************************
\section{The bouncing ball billiard}
\label{sec:DEFOFMODEL} 

\subsection{Equations of motion and control parameters of the model}

%*****************************************
\begin{figure}[h!]
\hbox{\hfill
\psfrag{A}{\hspace{-0.1cm}{\Large {$A$}}}
\psfrag{OM}{{\Large {$\omega$}}}
\psfrag{x}{{\Large {$x$}}}
\psfrag{y}{{\Large {$y$}}}
\psfrag{vx}{{\Large {$v_x$}}}
\psfrag{vy}{{\hspace{-0.4cm}\Large {$v_y$}}}
\psfrag{g}{\Huge $g$}
\psfrag{s}{\Large $s$}
\includegraphics[width=13cm,height=4.0cm]{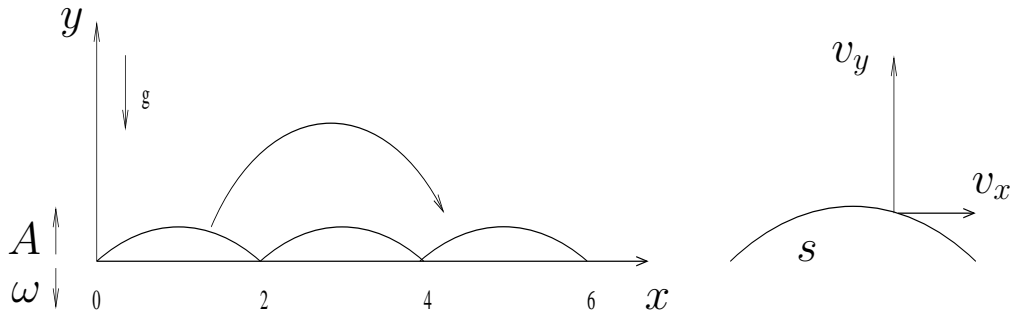}
%\scalebox{0.4}{\includegraphics{coord3.eps}}
\hfill}
\caption[small]{Illustration of the {\em bouncing ball billiard} studied here:
A moving point particle is colliding inelastically 
with circular scatterers that are 
periodically distributed on a line. Parallel to $y$ there acts an external
field with constant acceleration $g$. The floor of scatterers oscillates with
an amplitude $A$ and a frequency $\omega$.}
\label{fig:coordinates}
\end{figure}
%%****************************************

The bouncing ball billiard that we study in this paper, with the floor formed
by circular scatterers, is depicted in Fig.\ \ref{fig:coordinates}.

The equations of motion of this system are defined as follows: The particle
performs a free flight in the graviational field $g \parallel y$ between two
collisions. Correspondingly, its coordinates $(x^-_{n+1},y^-_{n+1})$ and
velocities $(v^-_{x\,n+1},v^-_{y\,n+1})$ at time $t_{n+1}$ immediately before
the $n+1$th collision and its coordinates $(x^+_n,y^+_n)$ and velocities
$(v^+_{x\,n},v^+_{y\,n})$ at time $t_n$ immediately after the $n$th collision
are related by the equations
\begin{eqnarray} 
x^-_{n+1} &=& x^+_{n} + v^+_{x\,n}(t_{n+1}-t_{n})  \\ 
y^-_{n+1} &=& y^+_{n} + v^+_{y\,n} (t_{n+1}-t_{n}) - g(t_{n+1}-t_{n})^2/2, \\
v^-_{x,\,n+1}&=& v^+_{x\,n}  \\ 
v^-_{y,\,n+1} &=& v^+_{y\,n} - g (t_{n+1}-t_{n})   \:.
\end{eqnarray}
At the collisions the change of the velocities is given by
\begin{eqnarray}
v^+_{\perp \,n}-v_{ci\perp\,n} &=&  \alpha
\,(v_{ci\perp n}-v^{-}_{\perp n})  \\
v^+_{\parallel \, n}-v_{ci\parallel \, n}
&=& 
\beta \,(v^-_{\parallel \, n}-v_{ci\parallel \, n}) \:,
\label{eq:horizrest}
\end{eqnarray} 
where $v_{ci}$ is the velocity of the corrugated floor. 
Here we distinguish between the different velocity 
components relative to the normal vector at the 
surface of the scatterers, where the scatterers are represented 
by the arcs of the circles forming the floor.  
$v_{\perp}$, $v_{\parallel}$ and $v_{ci\perp}$, 
$v_{ci\parallel}$ indicate the normal, respectively tangential 
components of the particle's, respectively the floor's velocity with respect 
to the surface at the scattering point. 
Correspondingly, we introduce two different
restitution coefficients $\alpha$ and $\beta$ that are perpendicular,
respectively tangential to the normal.

As in case of the standard bouncing ball problem we assume that the floor
oscillates sinusoidally, $y_{ci}=-A\,\sin(\omega\, t)$, where $A$ and $\omega$
are the amplitude respectively the frequency of the vibration, see
Fig.\ref{fig:coordinates}. Hence, the phase space is defined by the variables
$(t,x,y,v_x,v_y)$, where the time variable needs to be included because we
have a driven system described by nonautonomous differential equations.  If we
record only the collision events with the floor, as it is usually done in case
of the bouncing ball, the dimension of the phase space can be reduced
accordingly. For this purpose we use the position of the collision at the
circumference of a scatterer defined 
by the clockwise arclength at the impact $s$ and 
the horizontal and vertical velocities just after a collison, ${\bf
z}=(s,v_x^+,v_y^+)$. In such a Poincar\'e section the time-continuous flow is
equivalent to the time-discrete map
\bega
{\bf z}_{n+1} &=& f({\bf z}_n) \non\\
t_{n+1} &=& t_n+\tau ({\bf z}_n)  \\
l_{n+1} &=& l_{n}+a({\bf z}_n)  \non\:.
\ega    
called a {\em suspended flow} in Refs.\ \cite{Gas98,HaGa01,Gas96}. In the
second equation $\tau({\bf z})$ is the time of flight between two successive
collisions, and in the third equation $a({\bf z}_n)$ counts the number of
scatterers, or tiles, the particle has jumped over during a flight, $l_n$
being the tile where the particle has actually started at the $n$th collision,
hence $l_n$ is an integer. If the billiard is simple enough, these equations
can be derived analytically, such as in case of the periodic Lorentz gas
\cite{Gas98,Gas96} and in case of the completely elastic 
non-vibrating floor \cite{HaGa01}. This enormously simplifies the computer
simulations since the mapping has merely to be iterated numerically.  However,
in case of the bouncing ball billiard due to the vibrations we were not able
to derive the exact form of this mapping. Hence, the equations of motion were
solved via discretizing the parabolic flight between collisions in time and
trying to accurately determine the point of the collision with the surface  
by reducing the time lag of the iterations to $10^{-6}$s at collision 
events.

%************************************************************
\subsection{Choosing the parameters of the billiard}
\label{sec:parameters}

We choose the parameters of the billiard in Fig.\ref{fig:coordinates}
according to a possible experimental setup \cite{PrEgUr02}.  The extension of
each scatterer we determine to $d=2$mm. In our study we fix the value of
$\alpha$ and increase the value of $\Gamma$.  This can be done by tuning the
frequency $f=\omega/2\pi$ while keeping the amplitude constant at $A=0.1$mm.
The gravitational acceleration is taken to be $g=9.8$m/s$^2$.  The value of
the normal restitution coefficient is chosen to $\alpha=0.5$ such that the
major resonances as depicted in Fig.\ \ref{fig:at} do not overlap with each
other. Since our model has the geometry of a dispersing billiard, the arcs
induce a dynamical instability that somewhat counteracts the stability in
presence of a resonance. More importantly, the arcs make the system diffusive
by feeding energy that is vertically pumped into the system also into the
respective $x$-component of the velocity.

The main purpose of our model is to study the impact of dynamical correlations
on the deterministic diffusion coefficient, hence we would like to preserve
the mechanism related to the resonances of the bouncing ball problem as much
as possible.\footnote{We briefly remark that, by starting from the curved surface of Ref.\
\cite{HaGa01} and making the originally Hamiltonian dynamics dissipative,
first simulations seemed to indicate that the irregular structure of the
diffusion coefficient was rather immediately destroyed. That may be related to
the fact that certain periodic orbits associated to small islands of
stability, which indirectly determined this structure, are straightforwardly
eliminated by inclusion of dissipation. Furthermore, the oscillations of the
plate profoundly disturbed the original Hamiltonian dynamics. The precise
nature of the transition between the Hamiltonian case and the dissipative case
may be an interesting problem for further studies.}  Consequently, the radius
of the scatterers should be very large. Here we choose the radius $R$ of the
arcs and the corresponding curvature to $K=1/R=0.04$mm$^{-1}$. Note that,
because of the spatial periodicity, this billiard is of the form of a
two-particle problem with periodic boundary conditions. 
Since the radii of the 
scatterer and of the particle are simply additive, a periodic 
surface with respective radii could be realized experimentally 
by choosing a suitably large radius 
of the moving particle (which, of course, may have other side effects on which
we do not elaborate here).

At parameter values where no resonance is possible the particle may ``stick''
to the surface for some time, that is, it will land at a certain position and,
because of the friction, it will be relaunched at the next period of the
oscillation.  However, in case of fully elastic tangential scattering with
$\beta=1$, because of the spatial extension long trajectories may occur at
which a particle ``slides'' along the surface in one direction even if its
vertical velocity relative to the surface of the scatterer is zero. In order
to avoid these rather unphysical horizontal slides, we set the value of
$\beta$ smaller than one. Numerically it turned out that a horizontal
restitution of $\beta=0.99$ is already sufficient to eliminate all slides.  We
remark that this value is still considerably larger than the ones found in
experiments on inelastic particle collisions \cite{FoLuChAl94}, however, with
the choice of this value we avoid the vanishing of the  
diffusion coefficient at certain values of the frequency. 

Instead of using the dimensionless parameter $\Gamma$ we make our following
presentations in terms of the frequency $f$, since with a variation in steps
of $0.2$Hz it affects $\Gamma = A\,4\,\pi\,f^2/g$ only in the third digit
after the comma, consequently the frequency provides a much better scale. This
translates the position of the Arnold tongues in the bouncing ball problem at
$\alpha=0.5$, as shown in Fig.\ \ref{fig:at}, as follows: $1/1$-resonance:
$f\in [50.98, 61.56]$Hz; $2/1$-resonance: $[72.10,76.71]$Hz; $3/1$-resonance:
$[88.30,90.95]$Hz. The position of the tongues is also indicated in Fig.\
\ref{fig:difomega}. In the following, the diffusion coefficient is presented
in units of mm$^2$/s and the energy in kg\,mm$^2$/s$^2$.

%%***********************************************
%%*********************************************** 
\section{The frequency-dependent diffusion coefficient of the bouncing ball billiard} 
\label{sec:DIFCOEF} 

If we choose the set of parameters as explained in Sec.\ \ref{sec:parameters},
we expect that the dynamics of the spatially extended system is to a large
extent somewhat determined by the one of the bouncing ball problem.  However,
as was outlined in Sec.\ \ref{sec:BOBA}, for the bouncing ball problem time
and ensemble averages often strongly depend on the inital conditions, i.e.,
the system is nonergodic. In the bouncing ball billiard this deficiency is
eliminated by the defocusing character of the scatterers, i.e., the dynamics
is getting ergodic, except for some short frequency intervals at small
frequencies around the $1/1$-resonance. That is, the dynamics typically
evolves on only one or sometimes on two attractors.  The details related to
these phase space properties will be worked out in
Sec.\ref{sec:PHASEDIAGRAM}. Here we focus on the coarse structure of the
diffusion coefficient as related to regions of resonances.

\begin{figure}[h!]
{\hfill
\psfrag{D}{{\hspace{-1cm}$D$\,\,(mm$^2$/s)}}
\psfrag{f}{{\hspace{-0.5cm}$f\,\,\hbox{(Hz)}$}}
\includegraphics[width=13cm,height=10cm]{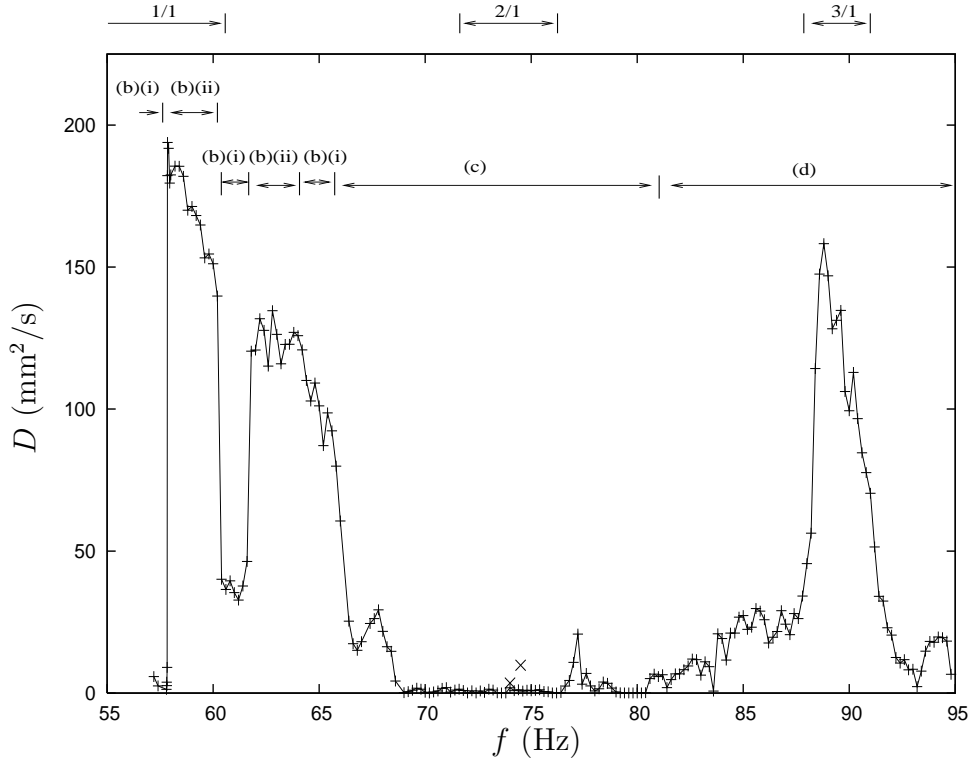}
\hfill}
\caption{The diffusion coefficient of the bouncing
ball billiard as a function of the vibration frequency of the corrugated
floor. The frequency range of the non-diffusive bouncing ball resonances, see
Fig.\ref{fig:at} at $\alpha=0.5$, are shown on top of the frame. They suggest
a strong impact of the resonances on the diffusion coefficient. However, the
fine structure of this curve appears to be more complex than merely explained
by the resonances.  Different dynamical regimes, as thoroughly discussed in
Sec.\ \ref{sec:PHASEDIAGRAM}, are indicated in this figure in form of
horizontal arrows. The two crosses $(\times)$ at the frequencies $74.0$Hz and
$74.5$Hz reveal the existence of the second resonance but were only obtained
by sampling a larger set of initial conditions. The standard deviation errors
are smaller than the magnitude of the symbols.}
\label{fig:difomega}
\end{figure}

For any frequency located in the interval of $f\in [49.77,95.0]$Hz the system
was found to evolve into a nonequilibrium steady state. This was checked in
terms of the average (kinetic, total) energy of the moving particle, which was
always approaching a constant value for large times.  The diffusion
coefficient in the horizontal direction was then obtained from the Einstein
formula
\begin{equation} 
D = \lim_{t\rightarrow \infty} \frac{<(x(t)-x(0))^2>}{2t}\:,
\label{eq:einstein} 
\end{equation} 
where the brackets denote an ensemble average over moving particles. The
diffusion coefficient was converging to a specific value for large enough
times. The size of the ensemble of points consisted of $10^3$ initial
conditions that uniformly filled the phase space region of $0\leq s\leq2.0004
\hbox{mm}$, $|v_x^+|\leq 40\hbox{mm/s}$, and $0<v_y^+
\leq40\hbox{mm/s}$.\footnote{For a considerable number of frequency values the
results for the diffusion coefficient have been cross-checked by sampling the
initial conditions for the ensemble from one long trajectory. That is, we take
as initial conditions coordinates from this trajectory after the dynamics has
evolved a random fraction of a maximal time that is considerably larger than
the time of typical transients. Within errorbars, the same results have been
obtained for the diffusion coefficient. However, this cross-check could not
easily be performed for frequency intervals marked by (b)(i) in Fig.\
\ref{fig:difomega} where the dynamics is nonergodic. In these cases the
ensemble was still the one described in Sec.\ \ref{sec:DIFCOEF}, apart from
the two points indicated in Fig.\ \ref{fig:difomega}, and the results may thus
depend on the choice of initial conditions if other attracting sets were not
sampled by the specific set of initial conditions.}

%************************************************
%*************************************************

The dependence of the diffusion coefficient on the frequency is depicted in
Fig.\ref{fig:difomega}. This figure clearly demonstrates that whenever the
value of the frequency reaches the different major Arnold tongues as discussed
in Sec.\ref{sec:BOBA}, the diffusion coefficient has considerably larger
values than elsewhere and consequently exhibits local maxima.

Above the frequency regime corresponding to the $1/1$ Arnold tongue, that is,
for $f\ge 70$Hz, the diffusion coefficient becomes very small but is still
different from zero in that it has a value around one or two.  This phenomenon
lasts until $f\simeq 80$Hz. Note that for a smaller tangential restitution
coefficient this region would collapse to $D=0$mm$^2$/s.
Fig.\ref{fig:difomega} shows that at the frequencies associated to the $1/1$
and to the $3/1$ resonances the diffusion coefficient is about two orders of
magnitude larger than the ones at $f\in[70,80]$Hz.

\begin{figure}[h!]
{\hfill
\psfrag{f}{\hspace{-0.5cm}{\Huge $f$\,\,(Hz)}}
\psfrag{E}{\hspace{-2.0cm}{\Huge $E$\,\,(kg mm$^2$/s$^2)$}}
%%\psfrag{Ekinx}{{\Huge $E_{kin,x}$}}
\hbox{
 \scalebox{0.4}{\rotatebox{270}{\includegraphics{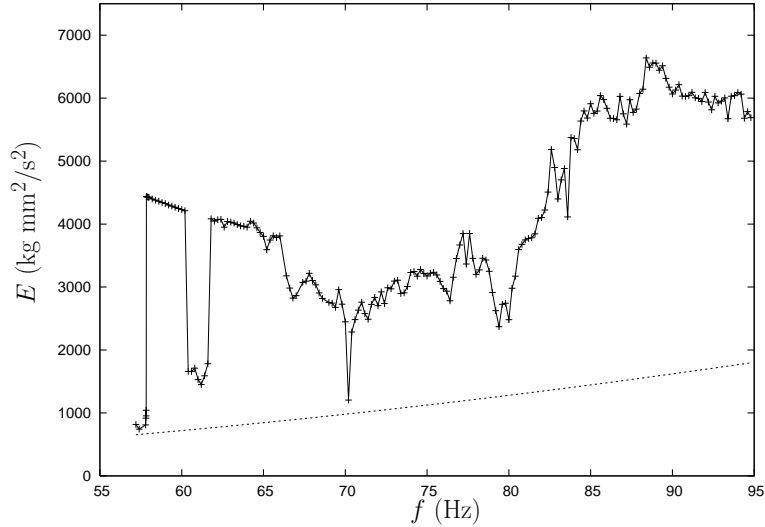}}}
 }
\hfill}
\caption[small]{The average total energy as a function of the vibration
frequency. The dotted line shows the possible minimal energy a particle can
have, which is the energy of a harmonic oscillator sticking to the surface, 
$A^2\omega^2/2$.}
\label{fig:energy1}
\end{figure}

\begin{figure}[h!]
{\hfill
\psfrag{f}{\hspace{-0.5cm}{\Huge $f$ \,\,(Hz)}}
\psfrag{Ekinx}{\hspace{-2.5cm}{\Huge $E_{kin,x}$\,\,(kg mm$^2$/s$^2)$ }}
\hbox{
\scalebox{0.4}{\rotatebox{270}{\includegraphics{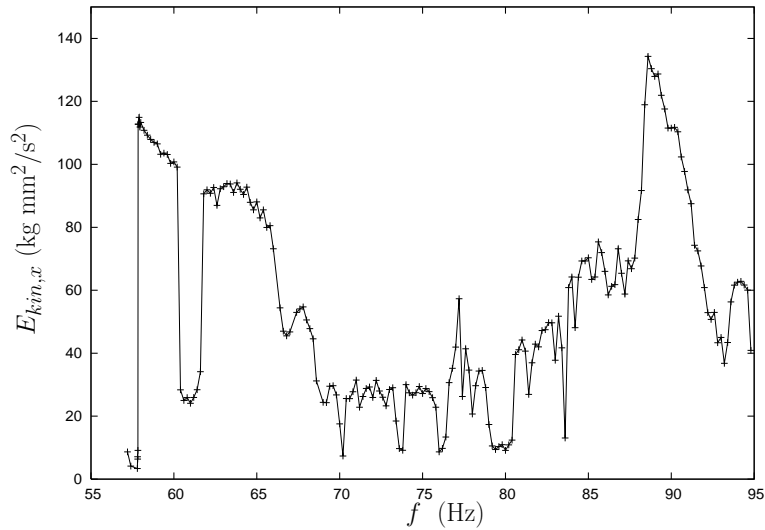}}}
}
\hfill}
\caption[small]{The average horizontal kinetic energy, that is,
$E_{kin,x}=v_x^2/2$, as a function of the vibration frequency.  One can see
that the structure of the kinetic energy closely follows the one of the
diffusion coefficient in Fig.\ \ref{fig:difomega}.}
\label{fig:kinen1}
\end{figure}

Irrespective of the specific frequency value, after a transient time the
average of the total energy converges to a constant value indicating the
existence of a nonequilibrium steady state. These values are presented in
Fig.\ \ref{fig:energy1} as a function of the frequency.  This curve resembles
very much the parameter-dependent diffusion coefficient. However, as may be
expected intuitively the value of the diffusion coefficient 
is even more closely
associated with the average kinetic energy related to the horizontal velocity
component $v_x^2/2$, see Fig.\ref{fig:kinen1}. Still, in detail the structure
of the diffusion coefficient cannot be trivially understood on the basis of
the frequency dependence of the (kinetic) energy only, or vice versa; see
Sec.\ \ref{sec:GRKU} for further details. It is aslo interesting to note here
that there is no equipartitioning of energy between the $x$ and the $y$
coordinate.  Fig.\ \ref{fig:energy1} furthermore includes a lower bound for
the energy of a moving particle, which presents the case when it sticks to the
surface and behaves like a simple harmonic oscillator. This approximation 
indicates an increase of the total energy, and of the corresponding 
horizontal kinetic energy, 
on a large scale which, however, appears to be largely
obscured by the superimposed oscillations of the energy.
%***************************************
%************************************** 

%***************************************
%***************************************

A further interesting feature of these three figures is that above $f\simeq
80$Hz particularly the diffusion coefficient, and to some extent also the
average energies, seem to experience a pronounced average increase on larger
scales. This property somewhat reminds of the phase transition-like behavior
discussed in Ref.\ \cite{FaTeVuVi99} where the average horizontal velocity of a
granular multilayer of particles in an asymmtric periodic potential shows a
sudden growth from zero to a nonzero value.

Finally, we would like to draw the attention to a particularly interesting
detail of these curves: Around $f\simeq 60$Hz in the first resonance there
appears to be an almost perfect linear decrease of the average total energy,
and correspondingly of the associated kinetic energy, whereas just in the same
region the diffusion coefficient exhibits some rather regular oscillations on
fine scales. One may argue that here the variation of the energy, as a
function of the frequency, provides a linear ``ramp'' that essentially reduces
the bouncing ball billiard to the system of Ref.\ \cite{HaGa01}, where the
diffusion coefficient was studied as a function of the energy as a control
parameter. Indeed, the diffusion coefficient of Ref.\ \cite{HaGa01} too
exhibited some very regular, apparently fractal oscillations under variation
of the energy, which qualitatively very much resemble the ones that appear in
the respective small region of Fig.\ref{fig:difomega}. We will get back to a
more refined analysis of the fine structure of the whole curve in Sect.\
\ref{sec:GRKU}.

%%************************************************************
%%************************************************************
\section{The different dynamical regimes of the bouncing ball billiard} 
\label{sec:PHASEDIAGRAM}

Depending on the driving parameter $\Gamma$, or respectively on the frequency
$f$, the following dynamical regimes can be distinguished in the bouncing ball
billiard:

(a) When $\Gamma < 1$ the maximal acceleration of the floor is smaller than
the gravitational force, $A \omega^2 < g$, consequently once the particle
lands on the surface it will never leave it again. In this case the particle
is eventually located in one of the wedges that the arcs of two adjacent
circles form with each other. Hence, after a short transient period the
particle will be in phase with the surface and its relative velocity to the
floor will be zero.  For our parameters this happens up to the frequency
$f_0=49.77$Hz. For frequency values just above $f_0$ a particle can be
relaunched from the floor, however, because of the inelasticity of the
collisions and due to the height of the arcs typically the particle remains
trapped in the wedges of the billiard. Numerically we find that around
$f=57$Hz particles start to leave a wedge for the first time, as shown in
Fig.\ \ref{fig:difomega}, consequently we consider this value as the onset of
diffusion, despite the drastic increase that takes place right afterwards
around $f=58$Hz. 

\begin{figure}[h!]
{\hfill
\psfrag{s}{\hspace{-0.5cm}{\small $s$\,\,(mm)}}
\psfrag{vx}{\hspace{-1cm}{\small $v_x^+$\,\,(mm/s)}}
%%\psfrag{t}{{\small $t$\,\,(s) }}
%%\psfrag{E}{\small $E$}
\hbox{
\rotatebox{270}{\includegraphics[width=6.3cm,height=7.5cm]{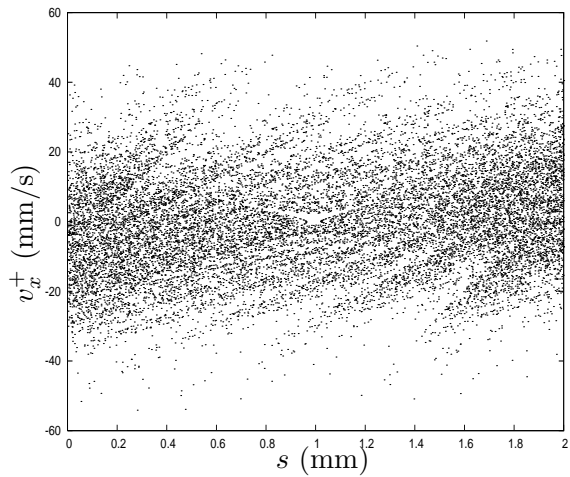}}
%%\hspace{0.2cm}
%%\rotatebox{270}{\includegraphics[width=6.3cm,height=5.5cm]{ener58.00.ps}}
}
\hfill}
\caption[small]{Projection of the phase space in the $1/1$ resonance where
the vertical dynamics is approximately in phase with the vibration frequency,
which is here $f=58$Hz. $s$ marks the clockwise position of the colliding
particle at the circumference of a scatterer and $v_x^+$ denotes the
horizontal velocity right after a collision.}
\label{fig:poincare1}
\end{figure}

(b) The most interesting dynamical regime appears to be associated with the
$1/1$-resonance of the bouncing ball. In case of a flat vibrating plate there
are two attractors at these frequencies, one showing this resonant behavior
and another one with particles that stick to the surface. In case of the
bouncing ball billiard this dynamical regime exhibits a complex scenario under
variation of the frequency, where due to the convexity of the scatterers also
ergodic motion becomes possible. This suggests to distinguish between two
different types of dynamics: (i) There are frequency regions where the system
has two coexisting attractors, one representing the $1/1$-resonance, and a
second one showing an intermittent-like behavior consisting of long periods of
stick and slip, or ``creepy'' motion.  This nonergodic regime is located at
the frequency intervals $f\in[57.0,57.8]$Hz, $f\in[60.4,61.8]$Hz and
$f\in[63.4,65.6]$Hz, see Fig.\ \ref{fig:difomega}.  (ii) The other type of
behavior is represented by frequencies at which there is only one attractor
with all inital conditions leading to the same phase locking. These frequency
intervals are at $f\in[58.0,60.2]$Hz and at $f\in [62.0,63.2]$Hz.  It is
remarkable that whenever particles are in resonance with the plate the motion
becomes regular only in the vertical direction.  In this direction we then
have a $1/1$ phase locked dynamics similar to the respective resonance of the
bouning ball. However, in the horizontal direction we still find a highly
irregular behavior as quantified by the diffusion coefficient.  In order to
illustrate this dynamics we look at a projection of the phase space at the
collisions of a particle with the floor. Because of the resonance the vertical
velocity has approximately the same value at the collision such that we almost
have a Poincar\'e section of the dynamics. Thus we only consider the position
$s$ of the colliding particle on the circumference of a scatterer and the
horizontal velocity $v_x$ right after a collision, see Fig.\ \ref{fig:poincare1}.

%%************************************************* 
%%*************************************************

\begin{figure}[h!]
{\hfill 
\psfrag{vx}{\hspace{-1cm} {\small $v_x^+$\,\,(mm/s)}}
\psfrag{s}{\hspace{-0.5cm} {\small $s$\,\,(mm)}}
%%\psfrag{t}{{\small $t$}}
%%\psfrag{E}{\small E}
\hbox{ 
\rotatebox{270}{\includegraphics[width=6.6cm,height=8.5cm]{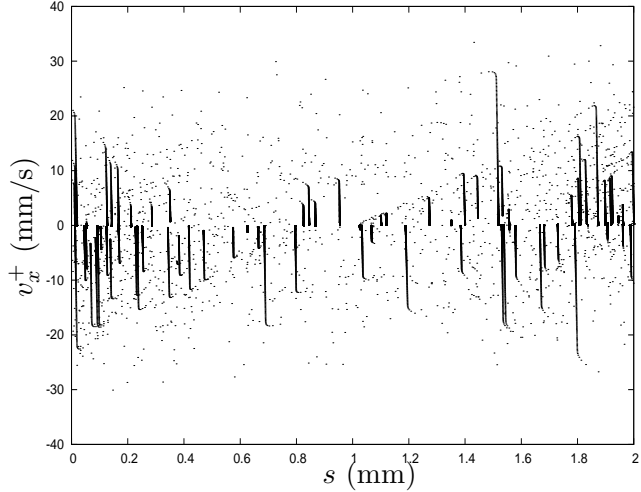}}
%\hspace{0.5cm}
%\rotatebox{270}{\includegraphics[width=6.3cm,height=5.5cm]{enf=85.20.ps}
}
\hfill}
\caption[small]{Projection of the phase space at the frequency of $f=85.2$Hz 
representing a dynamical regime where particles ``creep'' along the
surface. These creeps are visible in form of the almost vertical lines
indicating that a particle performs long sequences of correlated jumps nearly
at the same position with smaller and smaller horizontal velocities. The
coordinates are the same as in Fig.\ \ref{fig:poincare1}.}
\label{fig:birkf85-4}
\end{figure}

In this figure one can detect traces of a separatrix structure around the
unstable fixed point at $s=1,v_x^+=0$ corresponding to the two cases where a
particle can leave a wedge, i.e., the trajectory goes over $s=1$, or it cannot,
i.e., the particle does not approach the peak of an obsctacle or simply turns
back in the neigbourhood of $s=1$. In addition, there are interesting parts of
the phase space outside the separatrix in the regions of $s<1$, $v_x^+ > 0$,
respectively $s>1$, $v_x^+ < 0$. Here particles go "uphill" by generating a
fan-shaped structure in phase space showing that specific paths are chosen
according to which particles are allowed to leave a wedge, whereas other
possibilities are apparently forbidden. Yet there is no evidence that
particles creep along the surface of a scatterer in form of long sequences of
very tiny hops, cp.\ to Fig.\ \ref{fig:birkf85-4}, however, one may speculate
that the fan-shaped structure presents a precursor of such creeps. We
furthermore remark that for this figure the dynamics is ergodic and that the
average energy of a particle saturates after several hundred collisions such
that we get a good convergence at this frequency.

(c) Between the frequencies $f=65.8$Hz and $f=81.0$Hz the dynamics is
characterized by long creeps. Consequently, leaving the wedges takes a long
time and suppresses diffusion. In the neighbourhood of the frequency $74$Hz
the $2/1$-resonance could be detected, however, this resonance could only be
revealed by choosing a set of initial conditions that is considerably more
extended over the phase space than at the other frequencies, with initial
conditions of $v_y^+ > 100$m/s. The results indicated by the two specially
marked points in Fig.\ \ref{fig:difomega} at the frequencies $74$Hz and
$74.5$Hz show that the impact of this resonance on the diffusion coefficient
is, according to the respective measure of such orbits in the nonergodic phase
space, rather small. Note that, apart from this special case, we had no
indication for other non-ergodic behavior in regime (c). 

We furthermore note the existence of another small peak around $77$Hz that is
still close to the region of this resonance. 
Curiously, the following simple reasoning precisely identifies 
this frequency as a special 
point of the dynamics: Let us assume that a particle behaves similar to a
harmonic oscillator sticking to the surface. Then, as argued above, its
average vertical kinetic energy is equal to $E_y=A^2\omega^2/2$. However, for
diffusion this energy should be large enough such that a particle can pass an
obstacle, which is here of the height $h=0.02$mm. Consequently, if the
particle aquires a potential energy that is equal to $E_{pot}=g(A+h)$ it can
pass any scatterer even at the highest position of the vibrating plate. Of
course, in this simple reasoning we have disregarded any impact of the two
restitution coefficients as well as the fact that, additionally, some
horizontal kinetic energy is necessary in order to perform a diffusion
process. In any case, a respective calculation yields $f=77.23$Hz precisely
corresponding to the position of this small peak. One may furthermore
speculate that this argument is somewhat related to the onset of the global
increase of the diffusion coefficient and of the energy as discussed in Sec.\
\ref{sec:DIFCOEF}.

(d) For the next, in this study the last dynamical regime with
$f\in[81.1,95.0]$Hz Figs.\ \ref{fig:energy1} and \ref{fig:kinen1} demonstrate
a noticeable increase of the average total and horizontal kinetic energies
that appears to be reflected in a respective increase of the diffusion
coefficient on a coarse scale. Now the 
particle spends considerably more time for moving around than sticking to
the surface. Indeed, we find that the dynamics is ergodic such that, in
contrast to region (b), here both types of resonant and creepy 
motion are intimately linked to
each other, instead of appearing in coexisting attractors, or at different
frequencies. This type of dynamics is represented by the attractor of Fig.\
\ref{fig:birkf85-4}, where the coordinates $s$ and $v_x$ are shown at the
freqency $f=85.2$Hz.

%%*****************************************************
%%*****************************************************

In spite of the fact that particles creep from time to time along the
surface,\footnote{For the probability and the respective fraction of time of
sticking to the surface see \cite{LuMe93,Dev94}.} as we can clearly see in
the attractor of Fig.\ref{fig:birkf85-4}, the average energy of a particle
launched from randomly chosen initial conditions on this attractor saturates.
Consequenty, a system of noninteracting particles reaches a nonequilibrium
steady state in this regime, too.

%***********************************************
\begin{figure}[h!]
{\hfill
\psfrag{x}{\hspace{-0.5cm} {\small $x$\,\,(mm)}}
\psfrag{y}{\hspace{-0.5cm} {\small $y$\,\,(mm)}}
\psfrag{vy}{\hspace{-1cm} {\small $v_y^+$\,\,(mm/s)}}
\hbox{
\includegraphics[width=6cm,height=6.0cm]{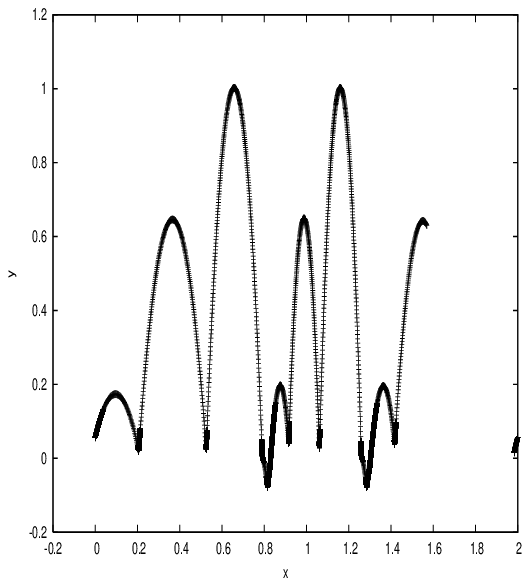}
\hspace{0.6cm}
\includegraphics[width=6.0cm,height=6.0cm]{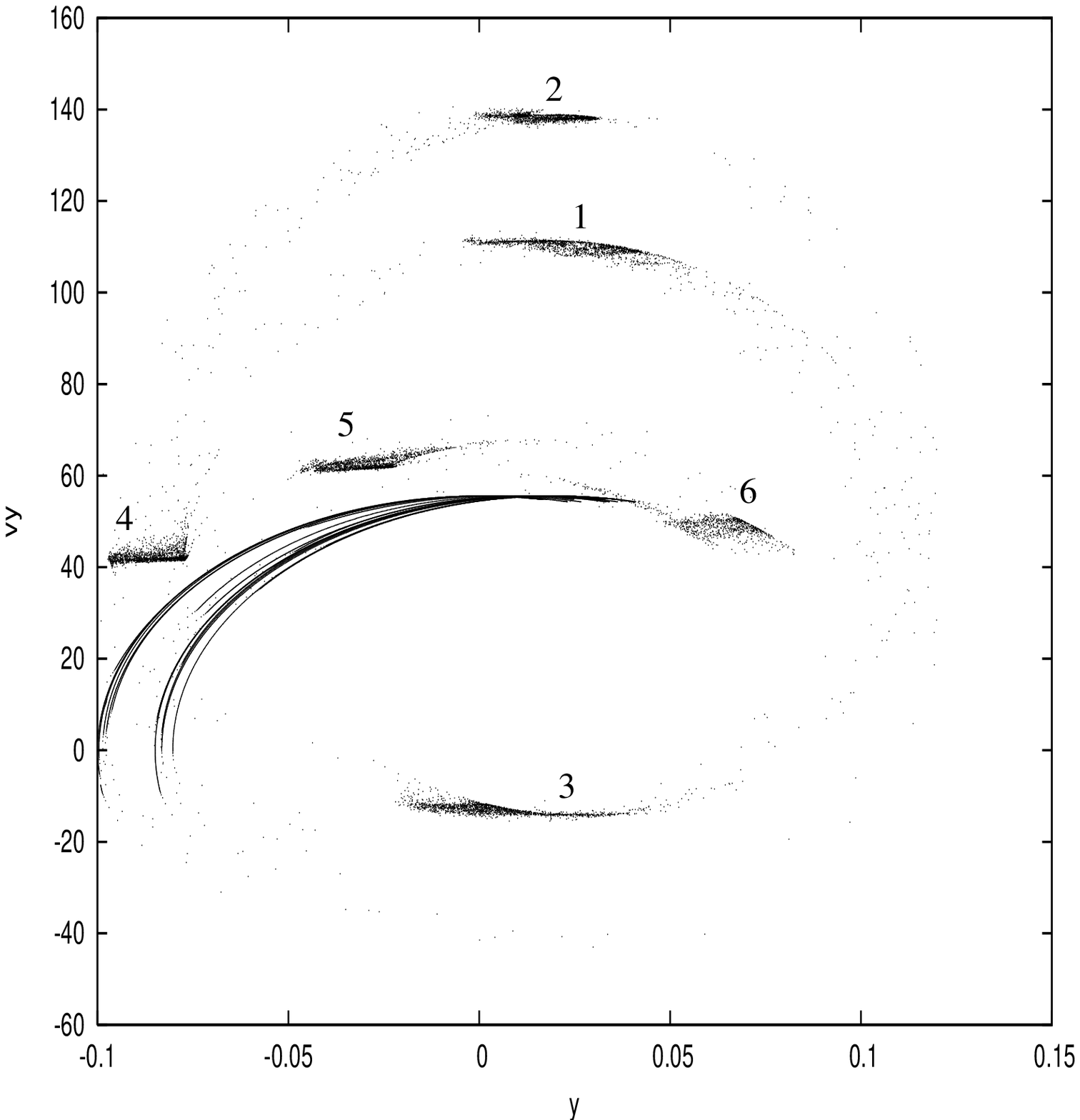}
}
\hfill}
\caption{A sequence of a particle's trajectory at $f=88.4$Hz (left) and the
corresponding attractor in the plane $(y,v_y^+)$ (right). These figures show
the deformation of the previous $3/1$-resonance into a much more complicated
quasi-periodic orbit.}
\label{fig:traj88-4} 
\end{figure} 
%************************************************

At the frequency interval of the $3/1$ resonance one can now find a
quasi-locked orbit, see Fig.\ref{fig:traj88-4}, that still induces a strong
maximum in the diffusion coefficient.  At first view this orbit indeed appears
to resemble a period $3$ orbit, however, closer inspection of the $y$
coordinate reveals that it may rather have bifurcated into portions that
resemble an orbit of period $6$. The right half of Fig.\ref{fig:traj88-4}
depicts these six regions that are visited one after the other.  However, an
even closer look reveals that in fact this is a period $11$ orbit: the
trajectory first follows the positions $1$ to $6$, but at any second sequence
it skips the position $6$ and goes instead directly from position $5$ to the
position $1$. Even more, sometimes the particle misses the last landing at
position $4$ and gets stuck at the surface.  It is then relaunched at the next
period, with the next bounce after the relaunch at position $5$, and the
periodic orbit starts again.

In summary, for all frequencies there appear to exist essentially four
different types of dynamical regimes: if the diffusion coefficient is very
large, the dynamics typically exhibits some type of periodicity that is
responsible for an enhancement of diffusion. Either this dynamics is strictly
associated to an original bouncing ball resonance, as in case of the regimes
(b)(ii), or the resonance still appears in a deformed way 
as in case of the largest peak in region (d). If the
diffusion coefficient is getting smaller there is an intermediate
regime in which two different attractors with resonances and creepy motion
coexist, and the dynamics is non-ergodic, see regimes (b)(i) 
and the small peak around $f=74$Hz. However, it is
also possible that the motion in this intermediate regime is intermittent-like
and ergodic, by alternating between localized creeps and long flights.
Finally, if the diffusion coefficient is getting closer to zero, there is
typically only sticky motion as discussed for region (c). Fig.\
\ref{fig:poincare1} exemplifies the regime of large diffusion coefficients,
whereas Fig.\ \ref{fig:birkf85-4} shows portions of typical creepy dynamics.
Note that, since we do not know about the position of any higher-order Arnold
tongues in the phase diagram Fig.\ \ref{fig:at}, we cannot possibly associate
smaller peaks in the diffusion coefficient to such higher-order tongues, but
we cannot exclude that this is possible. In any case, in the next section we
outline a simple approach that helps to explain the origin of the structure of
the diffusion coefficient particularly on smaller scales. 

%%******************************************************************
%%******************************************************************
\section{The fine scale of the coefficient: correlated random walk approximations}
\label{sec:GRKU}

As we explained in the previous section, the huge peaks in the diffusion
coefficient are associated with the major resonances of the bouncing ball
problem. However, first of all we have not yet clarified what large scale
functional dependence one would expect for the diffusion coefficient based on
stochastic theory. In addition, one observes a lot of irregularities on finer
scales in Fig.\ \ref{fig:difomega} whose origin needs to be commented upon. In
this section we work out some simple random walk arguments as well as
systematic refinements of it, which were called correlated random walk
approximations \cite{KlKo02}, in order to achieve a more detailed
understanding of Fig.\ \ref{fig:difomega}. 

\begin{figure}[h!]
{\hfill 
\psfrag{Drw}{\hspace{-1cm} {\Huge $D$\,\,(mm$^2$/s)}}
\psfrag{f}{\hspace{-0.5cm} {\Huge $f$\,\,(Hz)}} 
\scalebox{0.45}{\rotatebox{270}{\includegraphics{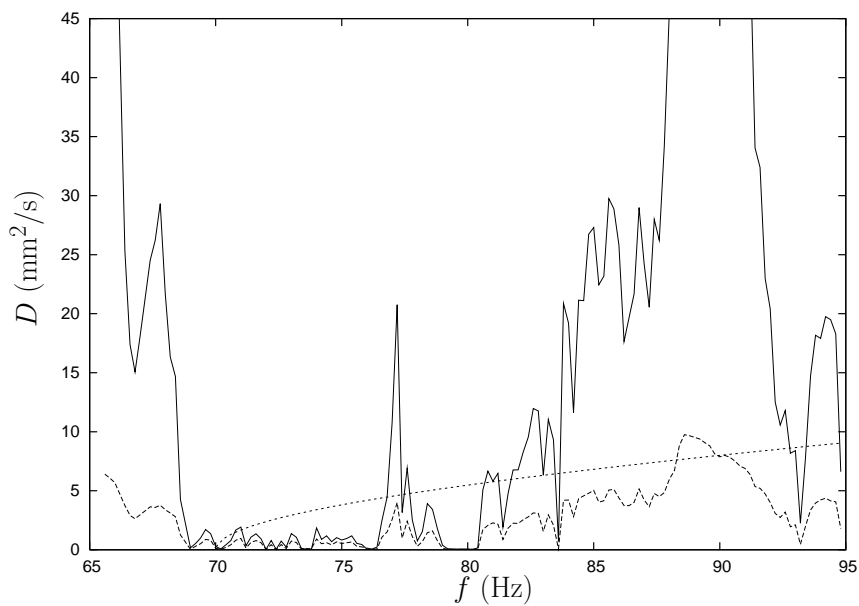}}}
\hfill}
\caption[small]{Numerical and analytical
evaluations of the random walk approximation Eq.\ (\ref{eq:machzwan}) for the
diffusion coefficient with frequencies $f > 65.6$Hz. The bold line shows again
the diffusion coefficient of Fig.\ \ref{fig:difomega}, the dotted monotonously
increasing curve represents the analytical formula Eq.\ (\ref{eq:analitic}),
and the dashed irregular curve close to zero depicts Eq.\ (\ref{eq:machzwan})
with the escape time computed numerically. The analytical approximation
indicates some increase of the diffusion coefficient on coarse scales, whereas
the numerical evaluation yields an irregular structure that is qualitatively
very similar to the one of the precise diffusion coefficient.}
\label{fig:grekub-all} 
\end{figure}

We first start with a simple approximation that is based on the picture of
diffusion as a random walk on the line. Let us assume that the wedges of the
billiard act as ``traps'' for a bouncing particle in which it stays for an
average time $\tau$. Let us furthermore assume that, after this time, the
particle escapes just to the neighbouring trap to the left or to the
right. Disregarding any correlations between subsequent jumps the random walk
diffusion coefficient is given by
\beq
D_{\rm rw}(f) = \frac{d^2}{2 \tau (f)} \:,
\label{eq:machzwan} 
\eeq 
where $d$ is the distance between two wedges and $\tau$ is the escape time of a
particle out of a wedge. For simple billiards such as Lorentz gases and
related models this formula can be worked out exactly
\cite{HKG02,KlKo02,MaZw83}. But here the dynamics is more complicated, hence
we first obtain $\tau$ from computer simulations by restricting ourselves to
frequency regions in which the dynamics is ergodic. The diffusion coefficient
Eq.\ (\ref{eq:machzwan}) resulting from this numerical input is shown in
Fig.\ref{fig:grekub-all}.

%********************************************************
%%*******************************************************

The irregularities resulting from $D_{\rm rw}$ are qualitatively very close
to the ones of the precise diffusion coefficient. This is easily understood by
approximating the average escape time to
\beq
\tau \simeq \frac{d}{<v_x>} \simeq \frac{d}{\sqrt{2E_x}} \:,
\label{eq:tau}
\eeq
hence the frequency dependence of the escape time must be very closely related
to the frequency dependence of the horizontal kinetic energy $E_x$ as shown in
Fig.\
\ref{fig:kinen1}, whose functional form was already observed to be close to
the one of the diffusion coefficient.

Eq.\ (\ref{eq:tau}) furthermore provides a convenient starting point for a
simple analytical approximation of the random walk diffusion coefficient Eq.\
(\ref{eq:machzwan}). In order to find an analytical estimate for $E_x$ we
start from the energy balance
\beq
E=E_x + E_y + E_{pot} \:,
\label{eq:balance}
\eeq
where $E$ is the average total energy and $E_y$ gives the average vertical
kinetic energy.  $E_{pot}=g\bar{y}$ is the average potential energy of the
particle.  We approximate the average height $\bar{y}$ with the amplitude of
the vibrations, $\bar{y} \simeq A$, and the
average total energy with the energy of the harmonic oscillator, 
$E \simeq A^2 \omega^2/2$.  The conjecture that the total energy of a
moving particle is roughly proportional to $\omega^2$ is also supported by a
simplified version of a vibrating plate \cite{WaHa95}. 
Finally, from computer simulations we approximate the relation
between horizontal and vertical kinetic energy to $E_y \simeq 19 E_x$
reflecting the fact that only a small percentage of the energy vertically
pumped into the system is transferred into the horizontal direction. This
appears to be due to the very shallow geometry of the corrugated floor and to
the action of the tangential friction. Taking all these expressions into
account we get from Eq.\ (\ref{eq:balance}) $2E_x\simeq A^2 \omega^2/20 -
gA/10$. Inserting this expression into Eq.\ (\ref{eq:tau}) and combining the
resulting formula for $\tau$ with Eq.\ (\ref{eq:machzwan}) we get a simple
analytical result for the random walk diffusion coefficient which is
\beq
D_a(f) \simeq \frac{d}{2} \sqrt{2E_x} 
\simeq \frac{d}{2} \sqrt{ \frac{A^2 \omega^2}{20} - \frac{gA}{10} }\:. 
\label{eq:analitic} 
\eeq 
This formula is depicted in Fig.\ \ref{fig:grekub-all}. After a discontinuous,
phase transition-like onset of diffusion it indicates some systematic but slow
increase of the diffusion coefficient on a coarse scale.  In the region of
$f\in[80,90]$ there is an analogy of this increase to the one obtained from
the numerical evaluation of Eq.\ (\ref{eq:machzwan}).

We now systematically refine the simple approximation Eq.\ (\ref{eq:machzwan})
in order to further explore the origin of the irregular structure of the
diffusion coefficient on fine scales. For this purpose we apply the scheme
proposed in Ref.\ \cite{KlKo02}, which suggests to approximate the precise
diffusion coefficient by adding higher-order terms to Eq.\ (\ref{eq:machzwan})
according to a Green-Kubo formula for diffusion.

%********************************************************
\begin{figure}[h!]
{\hfill 
\psfrag{D}{\hspace{-2.5cm} {\Huge $D_k \,\,\, \,\,\,D$\,\,(mm$^2$/s)}}
\psfrag{f}{\hspace{-0.5cm} {\Huge $f$ \,\,(Hz)}} 
\scalebox{0.45}{\rotatebox{270}{\includegraphics{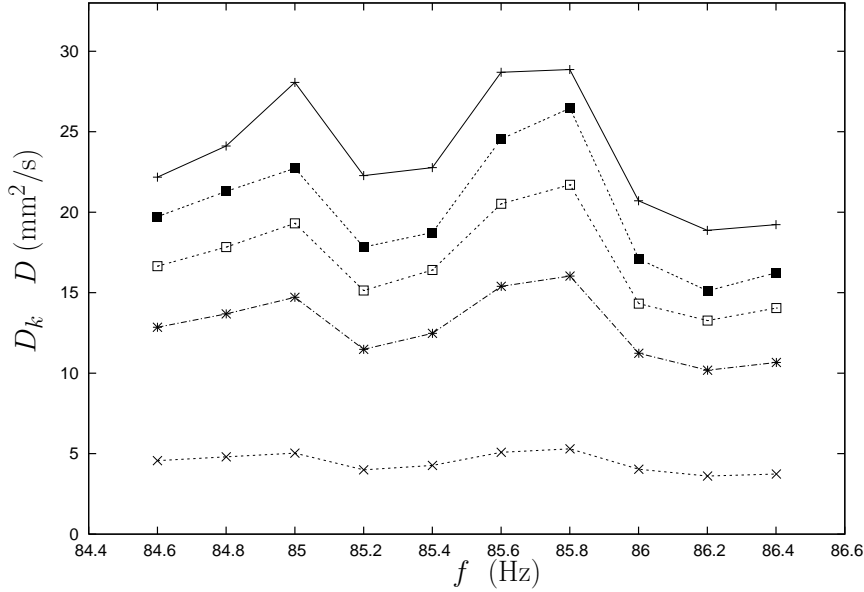}}}
\hfill}
\caption[small]{Magnification of the frequency interval $f\in [84.6,86.4]$Hz
in Fig.\ \ref{fig:difomega}. The bold line on top shows the respective
irregular diffusion coefficient on a fine scale, the lowest dotted line
depicts the corresponding random walk approximation. The other lines represent
a series of higher-order approximations refining this random walk and
converging to this curve.  These approximations $D_n(f)$ are obtained from
the truncated Green-Kubo formula Eq.\ (\ref{eq:gkapp}) for 
$n=0,2,4$ and $7$. The standard deviation errors are smaller than the 
magnitude of the symbols.}
\label{fig:grekub} 
\end{figure}
%%*******************************************************

Let us still assume that a particle moves with a frequency of jumps $1/\tau$
from traps to traps situated on a one-dimensional lattice at a distance
$d$. Starting from the Einstein relation Eq.\ (\ref{eq:einstein}) one can
prove that in the limit of infinitely many steps Eq.\ (\ref{eq:einstein}) is
equal to the sum
\beq 
D(f) = \frac{1}{\tau} \sum_{k=0}^{\infty} c_k 
< h(x_0) \cdot  h(x_k)>  \:,
\label{eq:greekubo1} 
\eeq 
which is a version of the Green-Kubo formula for an ensemble of particles
diffusing in a billiard \cite{KlKo02}. Here $ h(x_k)$ defines jumps of a
particle at position $x_k$ at the $k$th time step in terms of respective
lattice vectors, which here take the values $\pm d$. The first
coefficient is $c_0=1/2$, and for $k>0$ it is $c_k=1$.

We now truncate this series after the $n$th time step yielding the systematic
sequence of diffusion coefficient approximations 
\beq 
D_n(f) = \frac{d^2}{2\tau} + \frac{1}{\tau} \,\,\, \sum_{s_1 s_2 ...s_n} p(s_1
s_2\dots ) h \cdot h(s_1 s_2\dots)\:.
\label{eq:gkapp}
\eeq 
Here $s_1 s_2\dots$ denotes symbol sequences suitably identifying a
particle's trajectory passing through the traps \cite{KlKo02}. The averages in
Eq.\ (\ref{eq:greekubo1}) are now given in terms of the conditional
probabilities $p(s_1 s_2\dots)$ that characterize all these possible
trajectories.  The first term in Eq.\ (\ref{eq:gkapp}), that is, $D_0(f)$,
is again the random walk approximation Eq.\ (\ref{eq:machzwan}). 

Let us now sketch of how to evaluate Eq.\ (\ref{eq:gkapp}) to higher order.
Let the first step be in the positive (right) direction with $h(x_0)=d$. The
next jump is then defined by $ h(x_1)= d_{s_1}$, where $s_1$ can be either $f$
(forward) when the particle jumps to the right (in the same direction as the
first one) or $b$ (backward) when the particle jumps to the left, with $d_f=d$
and $d_b=-d$. If the dynamics is deterministic and highly correlated, a
particle starting from some tile may return to it with a probability that is
different from the backscattering probability of a random walk, which is
$1/2$. Hence, by feeding in the precise value for this probability as, e.g.,
computed from simulations one can suitably correct the random walk diffusion
coefficient $D_0(f)$ in first order.

In detail, the first approximation $D_1(f)$ at time step $2\tau$ reads
\beq
D_1 = D_0 + 2 D_0 (p_f-p_b) =D_0 +2 D_0 (1-2 p_b) 
\eeq
depending on the backscattering probability $p_b$, respectively the
forwardscattering probability $p_f$, only. In our case the value of $p_f$ is
considerably larger than the value of $p_b$, which is due to the fact that the
height of the arcs of the circles between two traps is very shallow, 
thus yielding a considerable contribution to $D_1$. 
In a similar way the second correction can
be evaluated to
\beq
D_2 = D_1 + 2 D_0(p_{ff}-p_{fb}+p_{bf}-p_{bb})\:.
\eeq
The probabilities associated to different symbol sequences were all computed
numerically, and for the frequency interval of $f\in[84.6,86.4]$Hz the
resulting approximations of increasingly higher order are shown in Fig.\
\ref{fig:grekub}. Evidently, the random walk diffusion coefficient $D_0(f)$ is
quantitatively much smaller than the numerically obtained value of the
diffusion coefficient. However, by adding more and more terms representing
repeated forward- and backwardscattering this series of approximations
converges to $D(f)$. Note that in order to obtain a good approximation one has
to go to relatively high order. That way, this figure exemplifies the impact of
strong dynamical correlations on the diffusion coefficient. Furthermore, this
convergence is obviously not monotonous in the frequency, with the higher
order approximations revealing irregularities on fine scales. This behavior is
well-known from related systems exhibiting deterministic diffusion
\cite{KlDo95,Kla96,Kla99,KlDe00,KlKo02,HKG02,HaGa01,KoKl02L} 
and typically indicates
the existence of a diffusion coefficient that is highly irregular as a
function of the parameter, or possibly even fractal. The origin of these
irregularities on finer scale may be attributed to the geometry of the
scatterers, which make the system topologically instable and induce a severe
pruning of trajectories that is reflected in the diffusion coefficient. In
other words, under parameter variation certain orbits may be forbidden, but
the nature of forbidden orbits may change in a very intricate way
\cite{Cvi01}.

In summary, we analyzed the frequency dependent diffusion coefficient on
different scales: On a coarse scale and for large enough frequencies the
diffusion coefficient appears to increase in qualitative agreement with a
simple analytical random walk approximation. However, this behavior is largely
suppressed by a series of huge peaks that we attribute to the impact of
resonances, which can be traced back to phase locking in the bouncing ball
problem. Finally, successive evaluations of the Green-Kubo formula of diffusion
give evidence for the existence of further irregularities on finer and finer
scales that indicate higher-order dynamical correlations, and one may
speculate that this curve is even of a fractal nature.

%%**************************************************
\section{Conclusions}
\label{sec:CONCLUSIONS}

The purpose of this work was to analyze deterministic diffusion in a novel
class of systems that we denoted as {\em bouncing ball billiards}. The
geometry of this class of models is very similar to the one of some well-known
chaotic Hamiltonian single-particle billiards such as Lorentz gases
\cite{Gas98,Do99} in consisting of a periodic array of scatterers together
with a point particle that collides with these obstacles. A particularly
interesting case was studied in Ref.\ \cite{HaGa01} composed of a periodically
corrugated one-dimensional floor in which a particle experiences a constant
vertical acceleration. Here evidence was given for the existence of a highly
irregular, possibly fractal diffusion coefficient, as it was already found to
exist in some very related, but somewhat simpler chaotic maps
\cite{KlDo95,Kla96,Kla99}.  Our goal was to move this mechanical system closer
to recent experiments on granular material, where similar geometries have been
used. Hence, we made the collisions inelastic by introducing normal and
tangential restitution coefficients. In this sense our model is somewhat
related to the inelastic Lorentz gas of Ref.\ \cite{KlRaNi00}. However, in
order to compensate for this loss of energy at the collisions the surface was
moving according to an oscillating driving force, just as it was done in
respective experiments. Finally, we composed the periodic surface of circular
scatterers that are very shallow. This way, our system is very close to the
well-known problem of a ball bouncing inelastically on a flat surface, which
is highly nonlinear and exhibits strong dynamical correlations in terms of
phase locking. Hence, this setup enabled us to study features of chaotic
diffusion in a specific granular one-particle system.

Our main numerical result is the existence of a highly irregular, possibly
fractal diffusion coefficient as a function of the driving frequency.  We
performed a detailed analysis in order to understand the complicated structure
of this curve and found that there are many different dynamical regimes
depending on the driving frequency. The most pronounced effects are due to
some previous regions of phase locking of the bouncing ball problem that still
manifest themselves in form of huge peaks in the diffusion
coefficient. Further features of our dynamics are parameter regions in which
the particle's dynamics may ``stick to'' and ``creep along'' the
surface. Although mostly associated to a very small value of the diffusion
coefficient, this dynamics may also come together with a quasi-periodic orbit
again exhibiting a huge peak at the position of a former bouncing ball
resonance. Finally, creeps and resonances may also coexist in form of some
nonergodic dynamics evolving on two different attractors. We furthermore
detected signs of a phase transition-like behavior at higher frequencies that,
to some extent, seemed to match to a simple analytical random walk
approximation. Refined approximations in form of correlated random walks,
based on the systematical numerical evaluation of a Green-Kubo formula of
diffusion, yielded evidence for the existence of further irregularities on
finer and finer scales pointing towards a possible fractal structure of the
frequency-dependent diffusion coefficient.

Due to the fact that both the bouncing ball and diffusion of granular
particles on periodic surfaces are systems that have been profoundly studied
in experiments \cite{FaTeVuVi99,PrEgUr02,Pi83,Pi85a,Pi85b,Pi88,Tu90}, we hope
that this paper stimulates further experimental work in order to verify this 
irregular frequency dependence of the diffusion coefficient. Of course we
would not expect to reveal any fractal curve in a real experiment. However, we
do believe that at least the largest peak is rather stable against random
perturbations, as appears to be corroborated by the studies in Refs.\
\cite{Kl02a,Kl02b}. Consequently, phenomena like the $1/1$-resonance may show up
in form of a local maximum of the diffusion coefficient as a function of the
frequency. Again, we emphasize at this point that without further studies
regarding the impact of perturbations on this billiard we cannot conclude to
which extent dynamical correlations such as the ones reported in this paper
may survive in case of a gas of interacting many particles. However, we note
that in Refs.\ \cite{LoCoGo99,NiBeCh00} the $1/1$-phase locking regime of the
bouncing ball was made responsible for the experimentally observed
\cite{LoCoGo99,OlUr98} phase transition between a quiescent amorphous and a
gaseous state in a monolayer of a granular gas.

Of course, more theoretical work needs to be done in order to predict more
details of possible experiments. There already exists an experimental setup
measuring diffusion of single granular particles in one dimension
\cite{FaTeVuVi99}, which is very close to our present model, however, the
respective channels are presently asymmetric. On the other hand, computer
simulations of a two-dimensional vibrating plate with circular scatterers
would be close to the experimental device of Ref.\ \cite{PrEgUr02}.
Furthermore, of course there exist more realistic models for inelastic
collisions than by using two constant restitution coefficients
\cite{FoLuChAl94}. For example, one may wish to include the possibility of
transfer of rotational energy of the moving particle at the collision.

One may also want to investigate the impact of a tilt on the bouncing ball
billiard making it into a ratchet-like device. On the basis of our studies and
related to the findings of Ref.\ \cite{GK01}, one may suspect that the current
may exhibit again a highly irregular, possibly fractal structure as a function
of control parameters reflecting the intrinsic properties of highly correlated
chaotic transport. First signs of such irregularities indeed appear to be
present in the numerical and experimental data of Ref.\ \cite{FaTeVuVi99}.  It
may also be elucidating to compute the spectrum of Lyapunov exponents and
possibly other dynamical systems quantities, as well as time-dependent
correlation functions and probability distributions of position and momentum
variables for such types of bouncing ball billiards\footnote{Note that
periodic fine structures of probability distributions that are analogous to
the ones reported in Ref.\ \cite{Wa96} were also detected in the chaotic maps
of Ref.\ \cite{Kla96}, hence we would conjecture that the bouncing ball
billiard shows the same type of distributions representing basic features of
deterministic dynamics.} and to check their characteristics under parameter
variation. Although simple reasoning suggests that the bouncing ball billiard
is chaotic whenever there is a resonance, in form of providing at least one
positive Lyapunov exponent, the features of the dynamical instability in
parameter regions of stickiness is far from clear.

Another crucial point is to learn more about the dynamics of the standard
bouncing ball problem with respect to the existence of higher-order
resonances. As was shown in Ref.\ \cite{LuMe90}, for an approximate bouncing
ball map in the limit of zero friction coefficient the complete set of tongues
forms a very complicated fractal Devil's staircase-like structure in the
parameter space. Unfortunately, for the general case up to now only the major
Arnold tongues have been calculated, and to find more tongues, to adequately
construct more details of the respective phase diagram, and to check in more
detail for the existence of bifurcations in the inelastically bouncing ball
problem appear to be open questions. However, in our case such information 
provided a road map in order to predict, and to understand, the irregular
structure of the frequency-dependent diffusion coefficient, hence a respective
analysis would be fundamental for further research along these lines.

Finally, we wish to remark that similar studies regarding the impact of phase
locking on diffusive properties have been performed in Refs.\
\cite{WA01,Ha02}. However, in both cases the systems under investigation were
very different from the granular model studied here: Ref.\ \cite{WA01}
outlined the impact of phase locking on the magnetoresistance in
antidot-lattices under constant electric and magnetic fields and equipped with
a bulk friction coefficient \cite{WA01}, whereas in Ref.\ \cite{Ha02} the
authors analyzed the impact of phase locking on the diffusion coefficient for
an overdamped stochastic Langevin equation with a ratchet-like asymmetric
periodic potential under the influence of a periodic tilt force. The same
Langevin equation, however, with a symmetric potential was used in Ref.\
\cite{Ha98} revealing again an enhancement of diffusion due to the matching of
an average waiting time of a particle with the driving frequency. This nice
paper as well as the very related studies of Ref.\ \cite{Fu75} 
point to a rather rich literature regarding the possible impact of
stochastic resonance-like phenomena on the diffusion coefficient. Of course,
the bouncing ball billiard too exhibits more important time scales, of which
the average escape time is an example, hence we would indeed conjecture that
there exist further resonances with the driving period in this model that may
lead to additional modulations of the diffusion coefficient. To find such
parameter regions further adds to the list of interesting open problems
related to the bouncing ball billiard. 

\section*{Acknowledgements} 
The authors wish to thank J.~Urbach for making them aware of important
literature regarding the bouncing ball problem. They are also grateful to him,
to A.~Prevost, to H.L.~Swinney and to I.~J\'anosi for remarks from the point
of view of the granular material. Interesting discussion with A.~Mechta and
A.~Pikovsky on synchronization and on the bouncing ball problem are gratefully
acknowledged. L.M. finally thanks H.~Kantz for 
discussions on the construction 
of steady states, and T.~T\'el for critical remarks on the manuscript.

%%***************************************************************

%***********************************************************


\begin{thebibliography}{00}

% \bibitem{label}
% Text of bibliographic item

% notes:
% \bibitem{label} \note

% subbibitems:
% \begin{subbibitems}{label}
% \bibitem{label1}
% \bibitem{label2}
% If there is a note, it should come last:
% \bibitem{label3} \note
% \end{subbibitems}

%\bibitem{}

\bibitem{Gas98}
P.~Gaspard, {\em Chaos, Schattering and Statistical Mechanics}, Cambridge
University Press, Cambridge, England, (1998).

\bibitem{Do99}
J.R.Dorfman, {\em An Introduction to Chaos on Nonequilibrium Statistical 
Mechanics}, Cambridge University Press, Cambridge (1999).

\bibitem{Vol02}
J.Vollmer, {\em Chaos, spatial extension, transport, and non-equilibrium
thermodynamics}, Phys. Rep. {\bf 372}, 131 (2002).

\bibitem{KlDo95}
R.Klages and J.R.Dorfman, Phys.\ Rev.\ Lett.\ {\bf 74}, 387 (1995).

\bibitem{Kla96}
R.~Klages, {{\it Deterministic Diffusion in One-Dimensional Chaotic Dynamical
Systems}, Wissenschaft \& Technik Verlag, Berlin (1996).}

\bibitem{Kla99}
R.~Klages, and J.R.~Dorfman,
\newblock {Phys. Rev. E} {\bf 59}, 5361 (1999).

\bibitem{KlDe00}
R.Klages and C.Dellago J.\ Stat.\ Phys.\ {\bf 101}, 145 (2000).

\bibitem{KlKo02}
R.Klages and N.Korabel J.\ Phys.\ A, {\bf 35}, 4823 (2002).

\bibitem{HKG02}
T.Harayama, R.Klages, and P.Gaspard, Phys. Rev. E {\bf 66}, 026211 (2002).

\bibitem{HaGa01} 
T.Harayama and P.Gaspard, Phys.\ Rev.\ E {\bf 64}, 036215 (2001).
 
\bibitem{KoKl02L}
N.Korabel and R.Klages 
Phys.\ Rev.\ Lett.\ {\bf 89}, 214102 (2002)

\bibitem{GK01} 
J.~Groeneveld and R.~Klages, J. Stat. Phys. {\bf 109}, 821 (2002).


\bibitem{LlNiRoMo95}
J.Lloyd, M.Niemeyer, L.Rondoni, and G.P.Morris, Chaos {\bf 5}, 536 (1995).


\bibitem{GaKl98}
P.Gaspard and R.Klages, Chaos {\bf 8}, 409 (1998).

\bibitem{Weis91}
D.~Weiss et~al., Phys. Rev. Lett. {\bf 66},  2790  (1991).


\bibitem{Weiss}
S.~Weiss {\it et al.}, Europhys. Lett. {\bf 51}, 499 (2000).

\bibitem{WA01}
J.Wiersig, K.-H.Ahn, Phys. Rev. Lett. {\bf 87}, 026803 (2001).

\bibitem{TKK02}
K.I.~ Tanimoto, T.~Kato, and K.~Nakamura, Phys. Rev. B {\bf 66}, 012507 (2002). 
\bibitem{FaTeVuVi99} 
Z.Farkas, P.Tegzes, A.Vukics, and T.Vicsek, Phys.\ Rev.\ E {\bf 60}, 7022
(1999). 

\bibitem{LoCoGo99} 
W.Losert, D.G.W.Cooper, and J.P.Gollub Phys.\ Rew.\ E, {\bf 59}, 5855 (1999).

\bibitem{PrEgUr02}  
A.Prevost, D.A.Egolf, and J.Urbach Phys.\ Rev.\ Lett. {\bf 89}, 084301 (2002).

\bibitem{Pi83} 
P.Pieranski, J.\ Physique {\bf 44}, 573 (1983).

\bibitem{Pi85a} 
P.Pieranski, Z.J.Kovalik, and M.Franaszek, J.\ Physique {\bf 46}, 681 (1985).

\bibitem{Pi85b} 
P.Pieranski and R.Bartolino, {J.\ Physique} {\bf 46}, 687 (1985).

\bibitem{Pi88} 
Z.J.Kovalik, M.Franaszek, and P.Pieranski, Phys.\ Rev.\ A {\bf 37} (1988)
4016.

\bibitem{Tu90}
N.B.Tufillaro, T.Abbott and, J.Reilly, {\em An experimental approach to
nonlinear dynamics and chaos}, Add.\ Wesley Publ., New York (1992).

\bibitem{GuHo83}
J. Guckenheimer and PJ Holmes {\em Nonlinear Oscillations, Dynamical Systems
and Bifurcations of Vector Fields}, Springer, New York, 5th edition (1997).

\bibitem{LL83}
A.~Lichtenberg and M.~Lieberman, Regular and Stochastic Motion, Springer, New
York (1983).

\bibitem{LuMe90}
A.Mechta and J.M.Luck, Phys.\ Rev.\ Lett. {\bf 65}, 393 (1990).

\bibitem{LuMe93}
J.M.Luck and A.Mechta, Phys.\ Rev.\ E {\bf 48}, 3988 (1993).

\bibitem{FeGrHuYo96} 
U.Feudel, C.Grebogi, B.R.Hunt, and J.A.Yorke, Phys.\ Rev.\ E {\bf 54}, 71
(1996).

\bibitem{Dev94}
P.Devillard, J.Phys. I {\bf 4}, 1003 (1994).

\bibitem{Gas96}
P.Gaspard, Phys. Rev. E {\bf 53}, 4379 (1996).

\bibitem{FoLuChAl94}
S.M.Foerster, M.Y.Louge, H.Chang, and K.Allia, Phys.Fluids {\bf 6}, 1108
(1994).
\bibitem{MaZw83} 
J.Machta and R.Zwanzig, Phys.\ Rev.\ Lett {\bf 50}, 1959 (1983).

\bibitem{WaHa95}
S. Warr and J.M. Huntley, Phys. Rev. E {\bf 52}, 5596 (1995).

\bibitem{Cvi01}
P. Cvitanovi\'c, R. Artuso, R. Mainieri, G. Tanner and G. Vattay, 
     {\em Classical and Quantum Chaos}, 
     {\tt www.nbi.dk/ChaosBook/}, Niels Bohr Institute (Copenhagen 2001) 

\bibitem{KlRaNi00}
R.Klages, K.Rateitschak and G.Nicolis, Phys.\ Rev.\ Lett.\ {\bf 84} 4268
(2002).

\bibitem{Kl02a}
R.Klages, Phys.\ Rev.\ E {\bf 65}, 055203(R) (2002).

\bibitem{Kl02b}
R.Klages, Europhys.\ Lett.\ {\bf 57}, 796 (2002).

\bibitem{NiBeCh00}
X.Nie, E.Ben-Naim and S.Y.Chen, Europhys.\ Lett.\ {\bf 51}, 679 (2000).

\bibitem{OlUr98}
J.S.Olafsen and J.S.Urbach, Phys.\ Rev.\ Lett.\ {\bf 98}, 4369 (1998).

\bibitem{Wa96}
S.Warr, W.Cooke, R.C.Ball, and J.M.Huntley, Physica A {\bf 231}, 551 (1996).

\bibitem{Ha02}
D.Reguera, P.Reimann, P.H\"anggi and J.M.Rub\'{\i} Europhys.\ Lett.\ {\bf 57},
644 (2002).

\bibitem{Ha98}
M.Schreier, P.Reimann, P.H\"anggi and E.Pollak Europhys.\ Lett.\ {\bf 44},
416 (1998).
 
\bibitem{Fu75} 
P.Fulde, L.Pietronero, W.R.Schneider, and R.Str\"assler, Phys.\ Rev.\ Lett.\
{\bf 35}, 1776 (1975).




\end{thebibliography}
\end{document}